\documentclass[letterpaper, 10pt, conference]{ieeeconf}
\IEEEoverridecommandlockouts
\usepackage{cite}
\usepackage[para]{footmisc}
\usepackage{amsmath,amssymb,amsfonts}
\usepackage{algorithmic}
\usepackage{graphicx}
\graphicspath{{./img/}}
\usepackage{textcomp}
\usepackage[table,xcdraw]{xcolor}
\usepackage{diagbox}
\usepackage{colortbl}
\usepackage{todonotes}
\usepackage{multirow}
\usepackage{url}
\usepackage{nohyperref}
\usepackage[binary-units,per-mode=symbol,group-separator={,},group-minimum-digits={3}]{siunitx}
\usepackage{subfig}
\usepackage[nohyperlinks,nolist]{acronym} 
\usepackage[]{algorithm2e}
\usepackage{tikz}
\usepackage{tabularx}
\newcolumntype{Y}{>{\centering\arraybackslash}X}
\newcolumntype{L}{>{\arraybackslash}X}
\newcolumntype{R}{>{\raggedleft\arraybackslash}X}
\newcolumntype{C}[1]{>{\centering\arraybackslash}p{#1}}
\usepackage{booktabs}
\usepackage[absolute,showboxes]{textpos}
\usetikzlibrary{patterns}
\usetikzlibrary{spy}
\usepackage{pgfplots}
\pgfplotsset{compat=newest}
\usepgfplotslibrary{units}
\usepgfplotslibrary{groupplots}
\makeatletter
\pgfplotsset{
 unit code/.code 2 args=
   \begingroup
   \protected@edef\x{\endgroup\si{#2}}\x
}
\makeatother
\usetikzlibrary{matrix}

\definecolor{CoreGray}{HTML}{BFBFBF}
\definecolor{CoreBlack}{HTML}{333333}
\definecolor{CoreBlue}{HTML}{002E7D}
\definecolor{CoreGreen}{HTML}{6AAC8E}
\definecolor{CoreRed}{HTML}{C80000}
\definecolor{CoreYellow}{HTML}{E6AC00}
\definecolor{CoreWhite}{HTML}{FFFFFF}
\definecolor{CoreLightBlue}{HTML}{8BBDEB}
\colorlet{LightCoreGray}{CoreGray!20}
\colorlet{LightCoreBlack}{CoreBlack!20}
\colorlet{LightCoreGreen}{CoreGreen!30}
\colorlet{LightCoreRed}{CoreRed!20}
\colorlet{LightCoreYellow}{CoreYellow!20}
\colorlet{LightCoreWhite}{CoreWhite!20}

\usepackage[absolute,showboxes]{textpos}



\begin{document}

\title{\LARGE \bf
A Multilayered Security Infrastructure for\\
Connected Vehicles -- First Lessons from the Field*
}

\author{Timo H\"ackel$^{1}$, Philipp Meyer$^{1}$, Lukas Stahlbock$^{2}$, Falk Langer$^{2}$,\\Sebastian A. Eckhardt$^{2}$, Franz Korf$^{1}$, and Thomas C. Schmidt$^{1}$%
\thanks{*This work is funded by the German Federal Ministry of Education and Research (BMBF) within the SecVI project (secvi.inet.haw-hamburg.de).}%
\thanks{$^{1}$\textit{Dept. Comp. Science}, \textit{Hamburg University of Applied Sciences}, Germany
        \{timo.haeckel, philipp.meyer, franz.korf, t.schmidt\}@haw-hamburg.de}%
\thanks{$^{2}$\textit{IAV GmbH}, Berlin, Germany
        \{lukas.stahlbock, falk.langer, sebastian.alexander.eckhardt\}@iav.de}%
}

\maketitle

\setlength{\TPHorizModule}{\paperwidth}
\setlength{\TPVertModule}{\paperheight}
\TPMargin{5pt}
\begin{textblock}{0.8}(0.1,0.02)
     \noindent
     \footnotesize
     Presented at the BROAD workshop at the 2022 IEEE Intelligent Vehicles Symposium (IV) in Aachen, Germany.
\end{textblock}

\begin{abstract}
    Connected vehicles are vulnerable to manipulation and a 
    broad attack surface can be used to intrude in-vehicle networks from anywhere on earth.
    In this work, we present an integrated security infrastructure comprising network protection, monitoring, incident management, and counteractions, which we built into a  prototype based on a production car. 
    Our vehicle implements a \acl*{SDN} Ethernet backbone to restrict communication routes, network anomaly detection to make misbehavior evident, virtual controller functions to enable agile countermeasures, and an automotive cloud defense center to analyse and manage incidents on vehicle fleets. 
    We present first measurements and lessons learned from operating the prototype: 
    many network attacks can be prevented through software-defined access control in the backbone;
    anomaly detection can reliably detect misbehavior but needs to improve on false positive rate;
    controller virtualization needs tailored frameworks to meet in-car requirements;
    and cloud defence enables fleet management and advanced countermeasures.
    Our findings indicate attack mitigation times in the vehicle from \SIrange{257}{328}{\milli\second} and from \SIrange{2168}{2713}{\milli\second} traversing the cloud.
\end{abstract}

%%%% 	Acronyms    %%%%
% !TEX root = ../main.tex
\begin{acronym}
	% A
	\acro{ACC}[ACC]{Adaptive Cruise Control}
	\acro{ACDC}[ACDC]{Automotive Cyber Defense Center}
	\acro{ACL}[ACL]{Access Control List}
	\acro{ADS}[ADS]{Anomaly Detection System}
	\acroplural{ADS}[ADSs]{Anomaly Detection Systems}
	\acro{ADAS}[ADAS]{Advanced Driver Assistance Systems}
	\acro{API}[API]{Application Programming Interface}
	\acro{AVB}[AVB]{Audio Video Bridging}
	\acro{ARP}[ARP]{Address Resolution Protocol}
	% B
	\acro{BE}[BE]{Best-Effort}
	% C
	\acro{CAN}[CAN]{Controller Area Network}
	\acro{CBM}[CBM]{Credit Based Metering}
	\acro{CBS}[CBS]{Credit Based Shaping}
	\acro{CNC}[CNC]{Central Network Controller}
	\acro{CMI}[CMI]{Class Measurement Interval}
	\acro{CoRE}[CoRE]{Communication over Realtime Ethernet}
	\acro{CT}[CT]{Cross Traffic}
	\acro{CM}[CM]{Communication Matrix}
	% D
	\acro{DoS}[DoS]{Denial of Service}
	\acro{DDoS}[DDoS]{Distributed Denial of Service}
	\acro{DPI}[DPI]{Deep Packet Inspection}
	% E
	\acro{ECU}[ECU]{Electronic Control Unit}
	\acroplural{ECU}[ECUs]{Electronic Control Units}
	% F
	\acro{FDTI}[FDTI]{Fault Detection Time Interval}
	\acro{FHTI}[FHTI]{Fault Handling Time Interval}
	\acro{FRTI}[FRTI]{Fault Reaction Time Interval}
	\acro{FTTI}[FTTI]{Fault Tolerant Time Interval}
	% G
	\acro{GCL}[GCL]{Gate Control List}
	% H
	\acro{HTTP}[HTTP]{Hypertext Transfer Protocol}
	\acro{HMI}[HMI]{Human-Machine Interface}
	\acro{HPC}[HPC]{High-Performance Controller}
	% I
	\acro{IA}[IA]{Industrial Automation}
	\acro{IDS}[IDS]{Intrusion Detection System}
	\acroplural{IDS}[IDSs]{Intrusion Detection Systems}
	\acro{IEEE}[IEEE]{Institute of Electrical and Electronics Engineers}
	\acro{IoT}[IoT]{Internet of Things}
	\acro{IP}[IP]{Internet Protocol}
	\acro{ICT}[ICT]{Information and Communication Technology}
	\acro{IVNg}[IVN]{In-Vehicle Networking}
	\acro{IVN}[IVN]{In-Vehicle Network}
	\acroplural{IVN}[IVNs]{In-Vehicle Networks}
	%J
	% L
	\acro{LIN}[LIN]{Local Interconnect Network}
	% M
	\acro{MOST}[MOST]{Media Oriented System Transport}
	% N
	\acro{NADS}[NADS]{Network Anomaly Detection System}
	\acroplural{NADS}[NADSs]{Network Anomaly Detection Systems}
	% O
	\acro{OEM}[OEM]{Original Equipment Manufacturer}
	\acro{OTA}[OTA]{Over-the-Air}
	%P
	\acro{P4}[P4]{Programming Protocol-independent Packet Processors}
	\acro{PCP}[PCP]{Priority Code Point}
	% R
	\acro{RC}[RC]{Rate-Constrained}
	\acro{REST}[ReST]{Representational State Transfer}
	\acro{RPC}[RPC]{Remote Procedure Call}
	% S
	\acro{SDN}[SDN]{Software-Defined Networking}
	\acro{SDN4CoRE}[SDN4CoRE]{Software-Defined Networking for Communication over Real-Time Ethernet}
	\acro{SIEM}[SIEM]{Security Information and Event Management}
	\acro{SOA}[SOA]{Service-Oriented Architecture}
	\acro{SOC}[SOC]{Security Operation Center}
	\acro{SOME/IP}[SOME/IP]{Scalable service-Oriented MiddlewarE over IP}
	\acro{SR}[SR]{Stream Reservation}
	\acro{SRP}[SRP]{Stream Reservation Protocol}
	\acro{SW}[SW]{Switch}
	\acroplural{SW}[SW]{Switches}
	% T
	\acro{TAS}[TAS]{Time-Aware Shaping}
	\acro{TCP}[TCP]{Transmission Control Protocol}
	\acro{TDMA}[TDMA]{Time Division Multiple Access}
	\acro{TSN}[TSN]{Time-Sensitive Networking}
	\acro{TSSDN}[TSSDN]{Time-Sensitive Software-Defined Networking}
	\acro{TT}[TT]{Time-Triggered}
	\acro{TTE}[TTE]{Time-Triggered Ethernet}
	% U
	\acro{UDP}[UDP]{User Datagram Protocol}
	\acro{UN}[UN]{United Nations}
	% Q
	\acro{QoS}[QoS]{Quality-of-Service}
	% V
	\acro{V2X}[V2X]{Vehicle-to-X}
	%W
	\acro{WS}[WS]{Web Services}
	% Z
	\acro{ZC}[ZC]{Zone Controller}

\end{acronym}

%%%%	Document    %%%%
%!TEX root = ../main.tex

\section{Introduction}%
\label{sec:introduction}

In future cars, we expect Ethernet-centric \acp{IVN}, which interconnect sensors, actuators, and further peripherals with clusters that bundle virtual functions and perform computationally intensive tasks.
Legacy devices will be integrated via gateways that comprise a zone topology~\cite{brkw-aeaet-17} -- all \acp{ECU} are connected to a zone controller in their physical vicinity.
Such Ethernet backbones form the basis for operations such as \ac{OTA} updates and new technologies such as \acl{ADAS}, and even autonomous driving.

Current \acp{IVN} are already vulnerable to manipulation by third parties. Attackers can exploit  multiple access interfaces to cars~\cite{cmkas-ceaas-11}, as has been shown by cyber-attacks in the field~\cite{mv-reupv-15}. Opening up \ac{V2X} communication with other vehicles, road side units, and the Internet will largely extend the attack surface and make connected cars a viable target for intruders. Protective measures at gateways and ECUs are required, but foremost the distributed vehicular system in total needs hardening by  a multi-sided security infrastructure. 

\begin{figure}[t]
    \vspace{4pt}
    \centering
    \includegraphics[width=1\linewidth, trim= 1.5cm 0.5cm 0cm 9.0cm, clip=true]{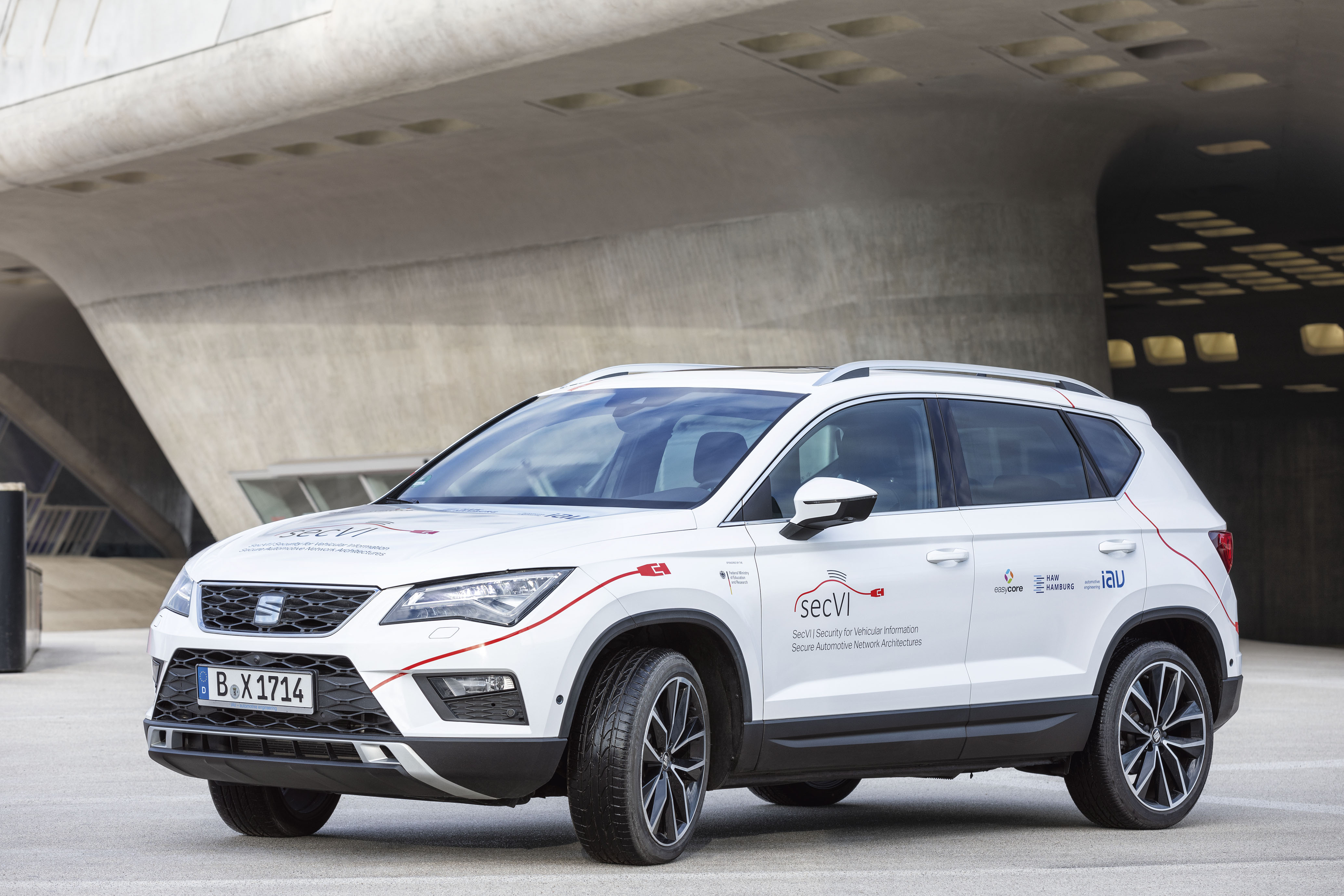}
    \caption{SecVI prototype car with security-enhanced network}
    \label{fig:prototype}
\end{figure}

In this paper, we present an integrated security infrastructure and its prototypic realization in a production vehicle (see Figure~\ref{fig:prototype}). Our \textit{Security for Vehicular Information} (SecVI) solution comprises network protection, monitoring, anomaly detection, incidence management, and selected countermeasures. We also report on first insights obtained while running the prototype for several months  after its initial demoing~\cite{mhlsd-dsivi-20}.

In the SecVI core, \acf{SDN}~\cite{mabpp-oeicn-08} is deployed to isolate flows that match the safety-critical communication of the vehicle. This increases robustness, security, and adaptability of the \ac{IVN}~\cite{hmg-rsarn-18, hmks-snsti-19, hhlng-saeea-20}.
A \acf{NADS} uses machine learning to fingerprint regular behavior, inspect network traffic, and detect misbehavior~\cite{rmwh-sadjr-18}.
A container orchestration framework manages applications on multiple computation nodes and dynamically adjusts their isolation~\cite{tvplj-cfcsoscof-19}, thereby enabling new incident response strategies. Externally, an \ac{ACDC} in the cloud performs monitoring and incident management for entire vehicle fleets~\cite{lss-eacdc-19}.

After reviewing background and the related work (Section~\ref{sec:background_and_related_work}), we introduce 
the security mechanisms integrated in our prototype vehicle (Section~\ref{sec:concept}). We show how the different security measures interact to protect the vehicle from cyber attacks. We further derive first insights from evaluations for all mechanisms (Section \ref{sec:eval}), including incident detection and reaction times for our prototype, which we compare with requirements from safety standards (Section~\ref{sec:response_time}). Finally, we conclude with an outlook on future work (Section~\ref{sec:conclusion_and_outlook}).

%!TEX root = ../main.tex

\section{Background and Related Work}%
\begin{figure}
    \vspace{4pt}
    \centering
    \includegraphics[width=1\linewidth, trim= 0.8cm 0.8cm 0.75cm 0.8cm, clip=true]{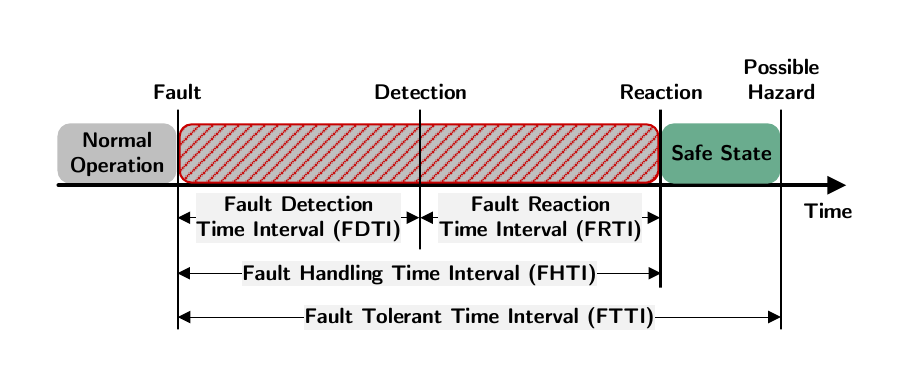}
    \caption{Fault Tolerant Time Interval according to ``Road vehicles -- Functional safety'' ISO 26262~\cite{iso-26262-1}}
    \label{fig:ftti}
\end{figure}

\label{sec:background_and_related_work}
Security incidents in vehicular components can quickly threaten the safety of the entire car. For safety, the ISO 26262 defines the \ac{FTTI} as the time a fault can be present in the system before a hazard occurs \cite{iso-26262-1}. The \ac{FTTI} depends on the corresponding safety goal and can range from a few milliseconds to some seconds. To prevent hazards, faults (e.g., security incidents) must be quickly detected and effective countermeasures (e.g., incidence responses) must follow. 
Figure \ref{fig:ftti} shows the relation between \ac{FDTI}, \ac{FRTI} and \ac{FTTI}.
In the following subsections, we explain the set of technologies used for security incident detection and reaction mechanisms implemented in the SecVI prototype car.

%!TEX root = ../main.tex
\subsection{Software-Defined Networking in Cars}
\label{subsec:background_software_defined_networking}
\acf{SDN}~\cite{mabpp-oeicn-08} separates the control logic (control plane) from the underlying switches that forward traffic on the data plane~\cite{krvra-sdncs-15}.
Simple network devices become programmable by a central \ac{SDN} controller using open standard protocols such as OpenFlow.
Controller applications implement the behavior of the network and enable, for example, routing protocols and access control.

In vehicles, \ac{SDN} can improve the safety and robustness~\cite{hmg-rsarn-18}, reduce the complexity and increase the adaptability of the vehicular E/E architecture~\cite{hhlng-saeea-20}.
In previous work, we integrated \ac{TSN} with \ac{SDN} to control in-vehicle communication without delay penalty for time-sensitive communication~\cite{hmks-snsti-19}.
\ac{SDN} also provides opportunities to improve network security~\cite{yd-sbcjr-21}.
This work builds on our approach to protect in-vehicle communication using the \ac{SDN} matching pipeline for precise access control~\cite{hsmks-sicfs-20}, and evaluates its performance while integrated in a prototype based on a production vehicle.

\subsection{Network Anomaly Detection for Cars}
\label{subsec:background_network_anomaly_detection}
\acp{IDS} are used to identify attacks on systems such as 
 hosts (H) or entire networks (N).
There are two IDS flavors.
(1) Signature detection systems, which can detect attack patterns saved in predefined attack signatures.
(2) \acp{ADS} use a predefined description of regular patterns and irregular patterns detected are potential attacks.
Since a large part of the communication within an IVN has a regular, predefined structure, our approach is a \acf{NADS} based on machine learning for fingerprinting normal behavior to inspect network (N) traffic.

Implementations of \ac{NADS} are diverse and in progressive development.
Various methods, systems, and tools are in use~\cite{bbk-nadms-14} and it is important to employ suitable datasets for their evaluation~\cite{rwslh-snbjr-19}.
To the best of our knowledge, there are no suitable datasets for evaluating \ac{NADS} in Ethernet based \acp{IVN}.
Network anomaly detection approaches for \acp{IVN} focus on traditional \ac{CAN} bus infrastructures~\cite{wmslk-sdvtm-17, rmwh-sadjr-18}.
Ethernet communication has different patterns, leading to different requirements for \ac{NADS}.
We present first lessons learned with a machine learning based Ethernet \ac{NADS} in a real car.

%!TEX root = ../main.tex

\subsection{Virtualized Automotive Microservices for Cars}
Application container use OS level virtualization to isolate processes and bundle software with required dependencies. For microservice compatible systems such as AUTOSAR Adaptive Platform containers can be used to integrate services. In automotive systems service orchestration enables vehicle operation in different modes such as autonomous driving and safe stop~\cite{kakwwnek-dsoahav-19}. In IT systems container orchestration frameworks are used to compose, manage and monitor a set of containers on multiple computation nodes~\cite{tvplj-cfcsoscof-19}. Using such a framework within the vehicle, allocation and isolation level of automotive services can be dynamically changed. We use these features for new security incident response strategies. In this work, the resource consumption of a container orchestration framework and reconfiguration time for \ac{SOME/IP} services on automotive reference development hardware is investigated.

\subsection{Cyber Defence Center for Cars}
Current vehicles have different online connections. Infotainment systems often include navigation with real time traffic, online maps, smartphone connections, as well as audio- and video streaming. Besides that online connections are also used for unlocking, summoning or updating the car. For future vehicle generations, even more online connections may be required to support Autonomous Driving Level 5.

To prevent drivers and environment from the risk of cyber-attacks, \ac{UN} and ISO/SAE are increasing their focus on vehicle cyber-security. As a result, the "\ac{UN} Task Force on Cyber security and \ac{OTA} issues (CS / OTA)" \cite{unece-wp29} and the norm on cyber-security \cite{iso-sae-21434} define a cyber-security management and monitoring system with incident response as a necessary part of a cyber-security system.

This results in the need for monitoring vehicles over lifetime with an \ac{ACDC} \cite{lss-eacdc-19} which is responsible for secure operation of vehicle fleets. An \ac{ACDC} collects information from vehicles security sensors (e.g. \ac{IDS} or logging) and analyzes these information to detect possible threats or incidences to a specific vehicle or a vehicle fleet. In case of an observed threat the \ac{ACDC} plans and executes necessary actions for an incidence response within the vehicle fleet. One can distinguish between three different stages: containment, eradication and recovery. To execute these stages, the vehicle needs to support different kinds of security actuators like reconfiguration and update mechanisms. Within this paper, an \ac{ACDC} infrastructure is set up to evaluate end-to-end detection and reaction times of security incidents.

%!TEX root = ../main.tex

\section{Secure Vehicle Prototype}
\label{sec:concept}

\begin{figure}
    \vspace{4pt}
    \centering
    \includegraphics[width=1\linewidth, trim= 0cm 0cm 0cm 5.0cm, clip=true]{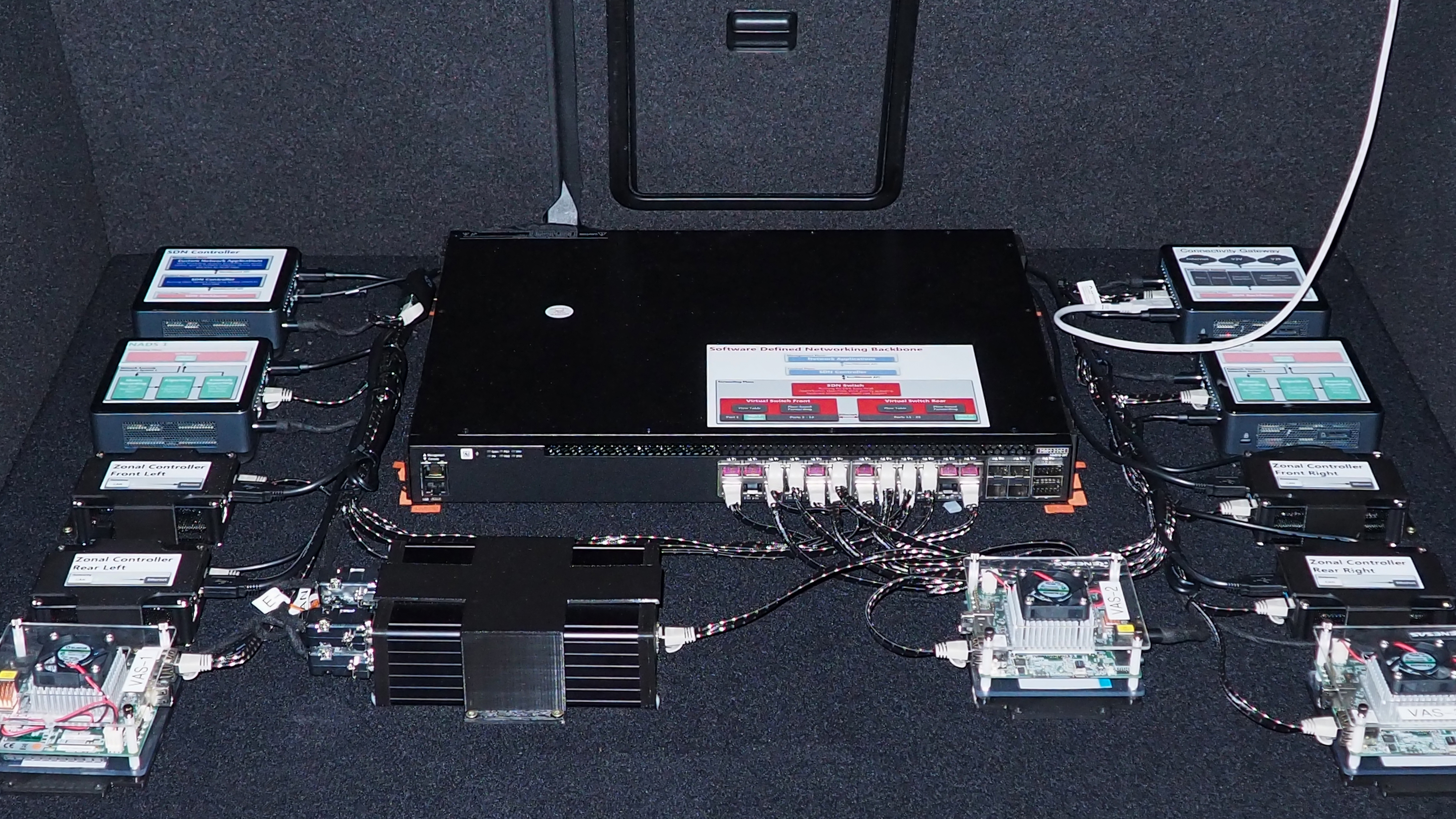}
    \caption{Installation in the trunk of the prototype vehicle}
    \label{fig:trunk}
    \vspace{-12pt}
\end{figure}

\begin{figure}
    \vspace{4pt}
    \centering
    \includegraphics[width=1\linewidth, trim= 0.8cm 0.6cm 0.8cm 0.6cm, clip=true]{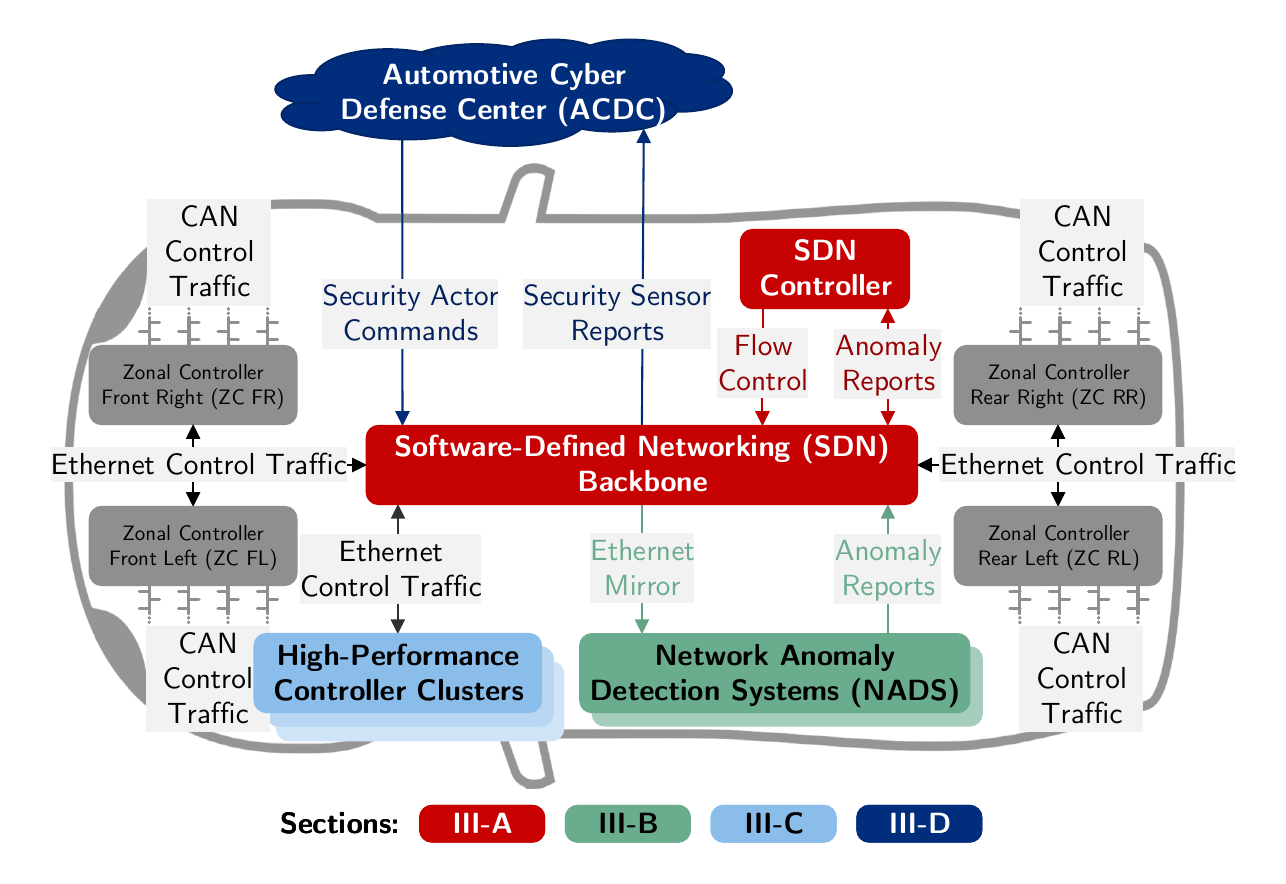}
    \caption{Prototype architecture of the security infrastructure}
    \label{fig:architecture}
    \vspace{-12pt}
\end{figure}

Our prototype builds upon a Seat Ateca shown in Figure~\ref{fig:prototype}.
We added an Ethernet zonal-architecture~\cite{brkw-aeaet-17} by splitting the original domain CAN busses into four zones (front left, front right, rear left, rear right).
In each zone, a \ac{ZC} acts as a gateway between an Ethernet backbone and the CAN devices that are in its physical vicinity.
We introduce additional native Ethernet communication such as high-resolution cameras sending raw camera images.
Figure~\ref{fig:trunk} shows our components installed in the trunk of the car and Figure~\ref{fig:architecture} shows a conceptual drawing of our security infrastructure.

%!TEX root = ../main.tex

\subsection{Hardened Software-Defined Backbone}
The Ethernet backbone consists of an OpenFlow-enabled switch, which is divided into two virtual Open vSwitch instances with programmable flow tables to steer packet forwarding. 
A flow table entry matches a subset of Layer~2 to Layer~4 header fields and performs actions like "discard", "forward", or "modify", if the traffic matches.
Mismatched packets are forwarded to the SDN controller for further inspection, which in turn decides how to forward them and can add, modify, and remove flow entries.
We use the ONOS \ac{SDN} controller with custom applications.

The \acp{ZC} repackage CAN messages into Ethernet frames and vice versa, which can be done on different layers with varying  impact on network security~\cite{hsmks-sicfs-20}. 
We use the \acf{SOME/IP} at the application layer, which is common with AUTOSAR. 
The CAN ID translates into the message ID of \ac{SOME/IP}, the data is embedded as payload.
Reserved UDP ports are used for \ac{SOME/IP} traffic.

The \acp{ZC} create a domain tunnel by sending all CAN messages of the same domain to one multicast destination address. All \acp{ZC} that communicate in this domain are subscribed to its multicast address.
Our flow entries for the CAN communication then match the source and destination IP addresses and UDP ports.

Flow based matching of \ac{SDN} combined with the well-known control communication of an \ac{IVN} enables very precise flow control. 
At start time of the IVN, a set of static flow rules is deployed that exactly match the native in-vehicle communication and only allow dedicated \acp{ZC} to speak to a CAN multicast tunnel.
The preconfigured static flow rules can act as a fallback configuration in an security incident or in case of a controller failure.
A second controller in hot standby mode could address the single point of failure, but would also increase deployment costs. 
The SDN controller can install new flows during runtime, which is regulated by an \acf{ACL} that for example can block switch ports, IP addresses, transport layer ports, or entire protocols.
\ac{ACL} violations are logged and reported to the ACDC.
If an anomaly is detected, the \ac{SDN} controller can reconfigure the network to ban certain flows or to disable all dynamic traffic.

\subsection{Ethernet Real-Time Anomaly Detection System}
To enable monitoring of all routes, the traffic from the two backbone switches is mirrored to two \acp{NADS}.
Each \ac{NADS} uses machine learning to fingerprint the behavior of individual communication streams.
Each stream consists of successive Ethernet frames identified by content relations like header fields.
One or multiple streams are assigned to an \ac{SDN} flow.
Violations of the learned stream behavior are reported to the \ac{SDN} controller and the \ac{ACDC}.

\begin{figure}[ht]
    \centering
    \includegraphics[width=1\linewidth, trim= 0.6cm 0.6cm 0.6cm 7.6cm, clip=true]{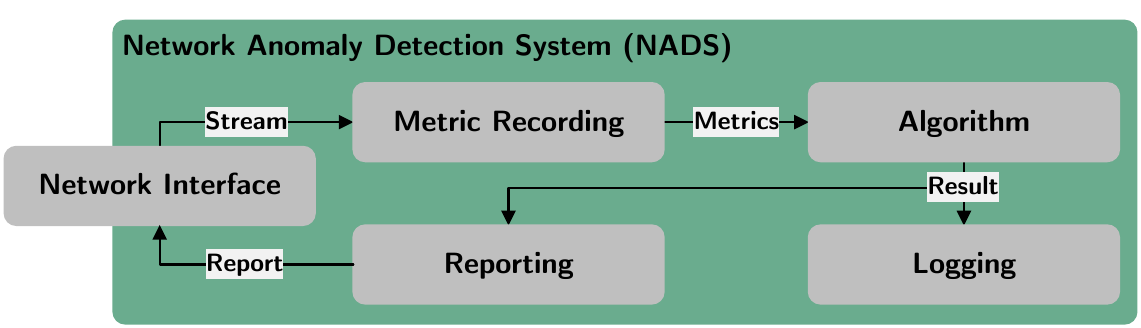}
    \caption{\acl{NADS} architecture}
    \label{fig:nads}
    \vspace{-4pt}
\end{figure}

Figure \ref{fig:nads} shows the architecture of the \ac{NADS} implementation in the prototype vehicle.
The \ac{NADS} processes all incoming Ethernet frames in four components: metric recording, algorithm, logging, and reporting.

Per stream, our \ac{NADS} processes multidimensional data and their relation in fixed intervals.
First, all raw Ethernet packets belonging to an observed stream are taken from a network interface and forwarded to the metric recording.
For every interval, this generates metrics from the raw data in normalized form (e.g., bandwidth, packet size, packet gap), which can be processed by the subsequent algorithm.
Different algorithms (e.g., clustering, classification, statistical) can be used in different states (e.g., learning, working).
The algorithm in turn extracts properties, which are then logged and reported back to the network as needed.
The implementation of all components is exchangeable and expandable and allows the comparison of different combinations of metrics and algorithms.

%!TEX root = ../main.tex

\subsection{Controller Cluster Service Orchestration}
The \ac{HPC} Cluster consists of three Renesas R-Car H3 Development Boards. For these nodes, a custom OS was built using \textit{yocto} (linux kernel: 5.4, K3S: 1.17.17+k3s1, containerd: 1.4.6, crun: 0.16, vsomeip: 3.16). We implemented K3S as container orchestration framework which is a lightweight Kubernetes distribution targeting embedded devices. A K3S cluster consists of several master or worker nodes~\cite{tvplj-cfcsoscof-19}. Workers are used to run the actual container workloads. Master nodes implement a control plane for management of containers and the system itself but can also run workloads. K3S works with PODs which are containers that bundle multiple coupled containers.

The orchestration framework is used to dynamically reallocate containerized \ac{SOME/IP} applications on \acp{HPC}. A \ac{SOME/IP} communication daemon runs on each \ac{HPC} to enable dynamic service discovery. Daemon and applications mount a shared path and create unix domain sockets for communication. This allows efficient inter-process communication although they run in different PODs. The \ac{SOME/IP} applications are scheduled as single-application PODs.

K3S provides a plugin-architecture for custom orchestration~\cite{tvplj-cfcsoscof-19}.
We implemented a module that determines the desired allocation of each \ac{SOME/IP} application based on the vehicle operation mode which is provided from the \ac{ACDC} depending on the systems security state.
In case of an observed threat, a reconfiguration may be used to disable potentially insecure applications or to isolate applications in order to maintain a secure vehicle operation.

\subsection{Automotive Cyber Defense Center}
The \ac{ACDC} is used for further processing of potential in-vehicle security incidents to decide which countermeasures can be initiated. It also enables aggregation of security related information of a vehicle fleet. Incident evaluation and definition of countermeasures may not always be an automated process but requires manual investigation of logs by security experts.

\begin{figure}[t]
    \vspace{4pt}
    \centering
    \includegraphics[width=1\linewidth, trim= 0.6cm 0.6cm 0.6cm 0.6cm, clip=true]{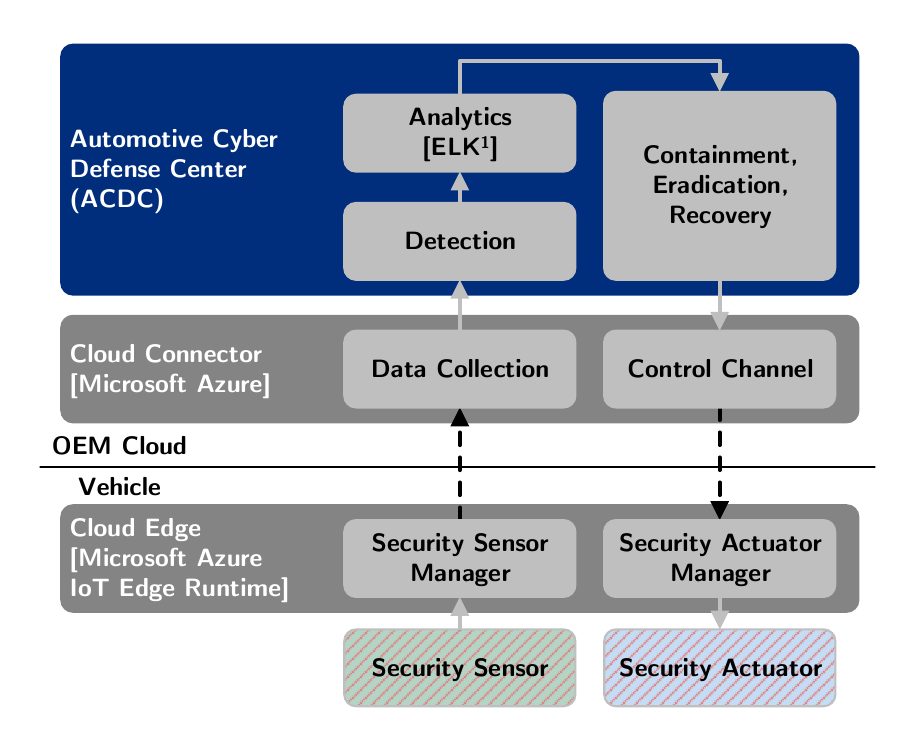}
    \caption{\acl{ACDC} architecture}
    \label{fig:acdc_modules}
    \vspace{-14pt}
\end{figure}

Figure \ref{fig:acdc_modules} shows the distributed \ac{ACDC} architecture.
To connect the vehicle and \ac{ACDC} we integrated Microsoft Azure IoT Edge Runtime within K3S on the \ac{HPC} cluster. The vehicle becomes a rolling edge device that can be monitored using existing infrastructures and tools. Within the K3S framework we deploy additional helper modules to setup the end-to-end security workflow.
(1) A security-sensor manager allows registration of in-vehicle security-sensors. It collects, aggregates and redirects security relevant information such as logs to Azure EventHub.
(2) A security-actuator manager waits for messages from the \ac{ACDC} backend and distributes the desired system state within the vehicle.
The security-sensor manager and security-actuator manager may also be used to implement local monitoring and local incident response strategies. For backend side evaluation of logs provided from the security-sensor manager the ELK\footnote{Elasticsearch, Logstash, Kibana (ELK)} stack was used to set up an Azure EventHub client. The backend itself is a virtual machine running in Microsoft Azure datacenter located in west europe. All other Azure related components are hosted in the same region.

%!TEX root = ../main.tex

\section{Lessons from the Prototype}
\label{sec:eval}
We now present first insights into the performance of the different security mechanisms from the deployment in the realistic hardware environment of the prototype vehicle. 

%!TEX root = ../main.tex

\subsection{SDN Flow Control in Attack Scenarios}

In connected cars, online capabilities are no longer limited to infotainment, but are also used for driving-relevant functions, such as up-to-date maps for autonomous driving.
The shared Ethernet backbone connects online devices to in-vehicle ECUs, opening up additional attack vectors.
Our SDN flow control restricts \ac{IVN} communication.
We discuss the benefits and limitations in typical attack scenarios.

Our network access control has the following configuration.
Static flows are preinstalled on both switches, 39 on one switch and 43 on the other.
These flows exactly match the arrival port, and layer~2 to layer~4 header fields of tunneled \ac{CAN} communication and other predefined communication. 
Traffic that does not match any installed flow is forwarded to the \ac{SDN} controller.
The \acf{ACL} in the controller is configured to strictly block most traffic, such as protocols like ICMP.
A whitelist on the controller describes the allowed communication for which new flow rules are dynamically installed on the network.

Attackers need access to the car to launch an attack.  
There are many interfaces, most of which are connected to larger \acp{ECU} such as the infotainment or an online gateway~\cite{cmkas-ceaas-11}.
Once attackers get past the first layer of defense, they can gain access to the \ac{IVN} backbone.
From there, they can launch untargeted attacks, such as probing the network, and targeted attacks on in-vehicle components.

To demonstrate the impact of our \ac{SDN} access control on untargeted attacks, we scan our \ac{IVN} with \textit{nmap}.
Hosts and their IP addresses could not be identified, as even pings are filtered.
TCP port scans do not return results for available services.
This is true even for ports to which there is a TCP connection from the source, since the flow rules match the source and destination ports and \textit{nmap} does not use the correct source port.
This shows that network access control is very effective in blocking untargeted attacks.

The static nature of \ac{IVN} also helps prevent attacks such as ARP and IP spoofing, since the addresses of the devices are known.
Other attacks such as \ac{DoS} or replay attacks can be prevented if the source is not allowed to access the specific flows.
However, such attacks are then forwarded by the switches to the controller and could overload it.
\ac{DoS} attacks on unknown flows, for example, are detected and prevented by \ac{ACL}, but unfortunately also pose a risk to \ac{DoS} on the controller.
This is a known problem in \ac{SDN} and protection mechanisms for the controller should be investigated in future work.

Targeted attacks require detailed knowledge about the vehicle. 
Attackers can target specific components, e.g., to gain control over the vehicle.
In previous work, we have shown that \ac{SDN} can control access to \ac{CAN} control traffic embedded in Ethernet flows, which reduces the \ac{IVN} attack surface~\cite{hsmks-sicfs-20}.
There are no flows from the online gateway to the zonal controllers, so no direct driving commands, such as braking, can be sent.
In future autonomous scenarios, the \ac{HPC} clusters will also use online services, e.g., for detailed up-to-date maps, and thus have communication paths to the online gateway.
Attacks on these valid flows, e.g. with forged information, are not prevented.

\begin{figure}[b]
    \centering
    \begin{tikzpicture}
        \begin{axis}[width=0.99\columnwidth, height=0.45\columnwidth,
            xbar stacked, xmin=0, xmax=1500,
            enlarge y limits=0.5,
            xlabel={Number of reports},
            symbolic y coords={Analytic, , NADS 2, NADS 1},
            xtick={250,500,...,1250},
            ytick={Analytic, NADS 2, NADS 1},
            yticklabel style={
                ,align=right
                ,inner sep=0
                ,xshift=-0.3em
            },
            ytick pos=left,
            nodes near coords,
            nodes near coords align={horizontal},
            ]
            \addplot[CoreGreen, fill=CoreGreen, every node near coord/.style={text=black}] coordinates {(955,NADS 1) (1274,NADS 2)};
            \addplot[CoreGray, fill=CoreGray, every node near coord/.style={text=black}, stack plots=false] coordinates {(1340,Analytic)};
        \end{axis}
    \end{tikzpicture}
    \vspace{-16pt}
    \caption{Number of true positive reports generated by \ac{NADS} 1 \& 2 video stream observation and the analytical expectation based on \ac{DoS} duration (\SI{134}{\second}) and assessment interval (\SI{0.1}{\second})}
    \label{fig:nads_reports}
\end{figure}
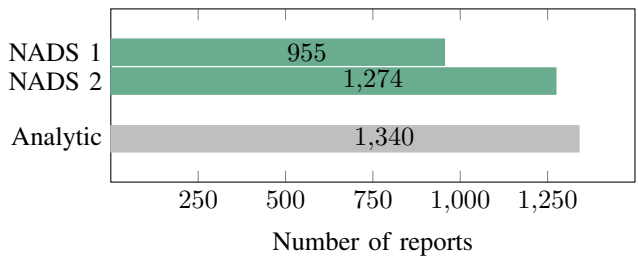

Our examples give first insights into what \ac{SDN} flow control can do to protect \ac{IVN}.
In general, communication is drastically limited and untargeted attacks on unknown flows are detected and blocked.
On the other hand, targeted attacks that use flows already installed in forwarding devices cannot be prevented by our access control mechanism.
Nevertheless, attacks are limited to the preinstalled and allowed flows and thus the attack surface is significantly reduced.

\subsection{Performance Analysis of Network Anomaly Detection}
Two individual \ac{NADS} are linked with the mirror ports of two \ac{SDN} backbone switches.
By monitoring both switches all possible traffic routes are observable.
This section will look at two selected flows traversing both switches.
In our prototype the behavior is fingerprinted with a mean shift clustering algorithm.
We plan on comparing further ML, deep learning, and heuristic algorithms~\cite{mhks-nadci-20}.
Each flow is tracked by an individual instance on \ac{NADS} 1 and 2:
\begin{itemize}
    \item \textbf{Control Traffic:}
    One original \ac{CAN} message stream encapsulated in SOME/IP$|$UDP$|$IP$|$Ethernet(802.3) to be forwarded over the backbone.
    The responsible \ac{NADS} entities use frame size and deviation from the message cycle as metrics X and Y.
    The algorithm assesses \SI{1}{\second} intervals.
    \item \textbf{Video Stream:}
    A video stream encoded in MPEG and encapsulated in UDP$|$IP$|$Ethernet(802.1Q) to be forwarded over the backbone (e.g camera to infotainment).
    The responsible \ac{NADS} entities use frame size and bandwidth as metrics X and Y.
    The algorithm assesses \SI{0.1}{\second} intervals.
\end{itemize}

Table \ref{tab:nads_fps} shows the proportion of true negatives and false positives assessment  intervals per \ac{NADS} instance during eight hours of operation with only regular traffic. 
\begin{table}[ht]
    \centering
    \caption{
        False positive assessment intervals during eight regular operating hours
        }
    \label{tab:nads_fps}
    \begin{tabularx}{\columnwidth}{c l Y Y}
        \toprule
        \textbf{NADS}    & \textbf{Flow}    & \textbf{Total} & \textbf{False Positives}\\
        \midrule
        \multirow{2}*{1} & Control Traffic  & 27666	         & 6\\
                         & Video Stream	    & 115701	     & 1\\
        \cmidrule(rl){1-4}
        \multirow{2}*{2} & Control Traffic  & 27608	         & 0\\
                         & Video Stream	    & 116182	     & 0\\
        \bottomrule
    \end{tabularx}
\end{table}

During those eight operating hours \ac{NADS} 2 reported no false positives.
There are 6 false positives for control traffic and 1 false positive for the video stream reported by \ac{NADS} 1.
This corresponds to a share of 0.02\% false positives for the control traffic and 0,0009\% false positives for the video stream.
This indicates, even if the proportion of false positives is very small those numbers are too high for this time span if alarms directly reflect in automated countermeasures or driver notifications.
Even with better performing algorithms and further reduced false positives a monthly driver notification or deactivated functions through countermeasures are not acceptable for customers.
Further local or cloud based filtering and report fusion is needed to decrease false positives and increase the quality of alarms.

The second measurement records the number of true positive alarms reported by the \ac{NADS} video stream instances during a plain \ac{DoS} attack.
The \ac{DoS} attack injects 10 million minimal UDP packets in \SI{134}{\second} (75076.6 packets per second) into the video stream flow.
Figure \ref{fig:nads_reports} shows the results in contrast with an analytical expectation.

\ac{NADS} 1 reports 955 alarms while \ac{NADS} 2 reports 1274 alarms.
The analytical expectation is calculated based on \ac{DoS} attack duration of \SI{134}{\second} and \ac{NADS} assessment interval of \SI{0.1}{\second}.
This results in an expectation of 1340 alarms.
From this follow 355 false negatives for \ac{NADS} 1 and 66 for \ac{NADS} 2.
This corresponds to a share of 5\% and 26\% false negatives.
This simple case indicates, even if the proportion of false negatives is high misbehavior is detectable.

The takeaway lesson of this exemplary implementation is that minimizing false positives is more crucial then reducing false negatives for deployment in future in-car \acp{NADS}.
Periodic false driver notifications or loss of functions through countermeasures are not acceptable for customers.

%!TEX root = ../main.tex
\subsection{Feasibility of Container Orchestration in Vehicles}

The general feasibility of existing container orchestration frameworks for dynamic system reconfiguration was investigated. We focused on two major aspects, resource consumption of the orchestration framework itself and the reconfiguration time of services. The R-Car H3 Board has a Cortex A53 quad-core, Cortex A57 quad-core and 4GB RAM. For all measurements, we disabled the A53 cores to have a homogenous computing architecture and to be closer to hardware used in current vehicle generation.

In order to evaluate the resource consumption we measured the CPU utilization and used memory using \textit{dstat} in two different workload scenarios. In the first workload scenario (base) we only had the K3S system modules running. In the second workload scenario (high) we deployed 50 pause-containers (rancher/pause:3.2) on each \ac{HPC}. These containers only run a sleep command and generate almost no additional CPU or memory stress. The measured overhead is therefore only caused by the K3S container management and monitoring facilities. For both workload scenarios, measurements were done on a master and worker node. To avoid additional delays caused by inter-node or module dependencies only one node was restarted at a time to monitor its startup behavior.

\begin{figure}
    \vspace{4pt}
    \centering
    \begin{tikzpicture}
        \begin{groupplot}[
            group style={group size=2 by 2,
            horizontal sep = 20pt, vertical sep = 20pt,},
            xmin=0,
            xmax=300,
            ymin=0,
            width=0.5\linewidth,
            no markers,
            /pgfplots/every axis plot/.append style={semithick},
            /pgfplots/legend image code/.code={
			\draw[mark repeat=2,mark phase=2] 
				plot coordinates {
					(0cm,0cm) 
					(0.3cm,0cm)
					(0.3cm,0cm)
				};
		    },
            yticklabel style={%
                ,text width=0.5cm
                ,align=right
                ,inner sep=0
                ,xshift=-0.3em
            }
            ]
            \nextgroupplot[ymax=100, ylabel={CPU util. [\%]}, title=base workload]
                \addplot [CoreBlue] table [x=t, y=base server cpu, col sep=semicolon, /pgf/number format/read comma as period] {data/k3s-framework.csv};
                \addplot [CoreLightBlue] table [x=t, y=base agent cpu, col sep=semicolon, /pgf/number format/read comma as period] {data/k3s-framework.csv};
                \legend{master, worker}
            \nextgroupplot[ymax=100, ytick=\empty, title=high workload]
                \addplot [CoreBlue] table [x=t, y=load server cpu, col sep=semicolon, /pgf/number format/read comma as period] {data/k3s-framework.csv};
                \addplot [CoreLightBlue] table [x=t, y=load agent cpu, col sep=semicolon, /pgf/number format/read comma as period] {data/k3s-framework.csv};
            \nextgroupplot[ymax=4100, ylabel={Memory usage}, y unit = \giga\byte, change y base, y SI prefix=kilo,xlabel={Time}, x unit = \second]
                \addplot [CoreBlue] table [x=t, y=base server mem, col sep=semicolon, /pgf/number format/read comma as period] {data/k3s-framework.csv};
                \addplot [CoreLightBlue] table [x=t, y=base agent mem, col sep=semicolon, /pgf/number format/read comma as period] {data/k3s-framework.csv};
            \nextgroupplot[ymax=4100, ytick=\empty, change y base, y SI prefix=kilo,xlabel={Time}, x unit = \second]
                \addplot [CoreBlue] table [x=t, y=load server mem, col sep=semicolon, /pgf/number format/read comma as period] {data/k3s-framework.csv};
                \addplot [CoreLightBlue] table [x=t, y=load agent mem, col sep=semicolon, /pgf/number format/read comma as period] {data/k3s-framework.csv};
        \end{groupplot}
    \end{tikzpicture}
    \caption{CPU utilization and memory usage while starting K3S on a master and worker node with base and high workloads
    }
    \label{fig:k3s-framework}
\end{figure}
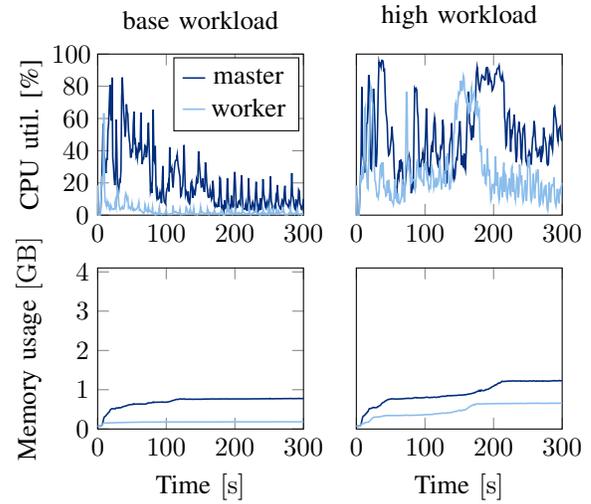

The average cpu utilization and memory usage of 10 independent system startups are shown in Figure \ref{fig:k3s-framework}. It can be seen that the K3S framework takes around 2 to 3 minutes for a worker node and 3 to 4 minutes for a master node to reach a constant cpu and memory consumption. Apart from the long starting time, in the high workload scenario the framework uses around 50/20\% CPU and 30/16\% RAM for a master/worker node.

\subsection{
    Findings from Operating an Automotive Defense Center
}
The collected incident reports and logs from in-vehicle security sensors in our prototype are forwarded to the \ac{ACDC} to detect attacks on one or more vehicles. 
Vehicles are nomadic devices with a non-permanent mobile network connection. 
Forwarding all vehicle information to the cloud would be ideal from an analytics perspective, but it also results in a lot of mobile data traffic and can add high costs to fleet operations.
A local in-vehicle component is needed to filter what information needs to be forwarded to the \ac{ACDC}. 
Within this component, sensor fusion of multiple security sensors can be performed to further reduce the false positive rates of individual security sensors, such as our \ac{NADS}. 
Filtering and sensor fusion can be implemented in the Security Sensor Manager.

Another challenge in experimenting with automotive specific algorithms to detect threats or incidents on vehicle fleet level is currently the very poor availability of realistic data. For IT systems there are sometimes data bases of log data with known attacks available. Equivalent data bases for vehicle fleets containing a ground truth are missing. To overcome this problem, we created a simulation that generates sample log data for \acp{IDS} in a vehicle fleet. The simulation data is concatenated in the \ac{ACDC} with the prototype data.

The edge computing based framework allows fast reaction times on security incidents even with a cloud involved. Some reactions can be triggered autonomously such as blocking a flow using \ac{SDN} or setting the vehicle in a predefined fail safe mode with a subset on available applications. Other reactions need deep manual inspection of logs and incident reports in order to find the root cause of an incident. The human factor increases the reaction time from seconds to days/weeks. If complex system updates are necessary to remedy bugs, the reaction time may even go up to months because of long software testing and certification processes. Our experiences with the prototype have shown that combined response strategies are more sufficient. First, an automated reaction to set the vehicle in a fail safe mode where the vehicle operation is secure (containment). Then the actual threat can be investigated (eradication). At last, a software update is provided (recovery).

\section{Incidence Response Time Measurements}
\label{sec:response_time}
The security mechanisms of the prototype can work together to effectively counter attacks. 
Analogous to the \acf{FTTI} from safety standards, we consider the following time intervals in our measurements:
\begin{itemize}
    \item The \textbf{\acf{FDTI}} from the beginning of the attack to its detection.
    \item The \textbf{\acf{FRTI}} from the incident report to the implementation of countermeasures.
    \item The \textbf{\acf{FHTI}} from the start of the attack until a safe state is reached.
\end{itemize}
Ideally, the \ac{FHTI} is the sum of the \ac{FDTI} and the \ac{FRTI}, but in reality additional transmission and processing delays have to be taken into account.
We distinguish between an in-vehicle \ac{FHTI}, where only \ac{IVN} components are used to detect and mitigate the attack, and a cloud \ac{FHTI}, where reports from in-vehicle components are sent to the \ac{ACDC} backend, which processes them and initiates countermeasures that are implemented in the vehicle.

\begin{figure}[t]
    \vspace{4pt}
    \centering
    \includegraphics[width=1\linewidth, trim= 0.8cm 0.6cm 0.8cm 0.6cm, clip=true]{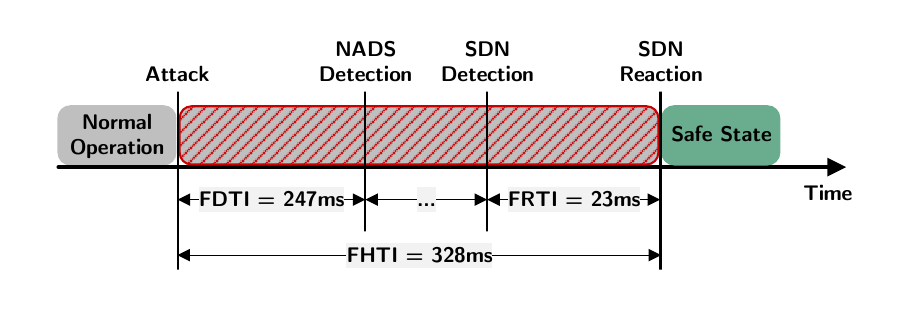}
    \caption{Maximum of in-vehicle incident response times}
    \label{fig:ftti_local}
\end{figure}

The \ac{FHTI} for the local incidence response is shown in Figure~\ref{fig:ftti_local}. 
The results include the min., max. and avg. for the time intervals from five independent measurement series, with the maximum shown in the figure. 
A \ac{DoS} attack is launched on the video stream from a camera to infotainment. 
We measure the time from the arrival of the first attack frame at the destination until no more attack frames arrive and the incident is resolved. 
Figure~\ref{fig:ftti_local} shows the maximum \ac{FHTI} of \SI{328}{\milli\second}. 
The measured minimum is \SI{257}{\milli\second} and the average is \SI{290}{\milli\second}. 
The \ac{FHTI} is composed of the \ac{NADS} detection (\ac{FDTI}) and the \ac{SDN} response time (\ac{FRTI}).

The measured \ac{FDTI} is the time between the first attack frame reaching the \ac{NADS} and the first outgoing anomaly report.
Figure~\ref{fig:ftti_local} shows the maximum \ac{FDTI} of \SI{247}{\milli\second}.
The measured minimum is \SI{80}{\milli\second} and the average is \SI{163}{\milli\second}.
The \ac{FDTI} is dependent on the \ac{NADS} assessment interval, which in the case of the video stream is \SI{100}{\milli\second}.
After each interval a report is generated if misbehavior is detected.
It must also be considered that the \ac{NADS} implementation is in prototype state.
Computation time would be significantly reduced in optimized deployments.

The \ac{NADS} report is transmitted over the backbone to the \ac{SDN} controller, with an communication latency added before the controller receives the \ac{NADS} report. 
When the \ac{SDN} controller detects the report, it looks for possible countermeasures. 
In this scenario, a countermeasure was configured to delete the malicious video stream flow from all network devices.

In our measurements, the controller sends these flow modification requests to both switches a minimum of \SI{7}{\milli\second}, an average of \SI{10}{\milli\second}, and a maximum of \SI{19}{\milli\second} after the first report arrives. 
Both switches acknowledge the flow removal a minimum of \SI{2}{\milli\second}, an average of \SI{5}{\milli\second}, and a maximum of \SI{16}{\milli\second} later.
The \ac{FRTI} is measured on the \ac{SDN} controller and starts when it receives the \ac{NADS} report and lasts until both switches have acknowledged all requested countermeasures.
Figure~\ref{fig:ftti_local} shows the maximum \ac{FRTI} of \SI{23}{\milli\second}.
The measured minimum is \SI{9}{\milli\second} and the average is \SI{15}{\milli\second}.

\begin{figure}[t]
    \vspace{4pt}
    \centering
    \includegraphics[width=1\linewidth, trim= 0.8cm 0.6cm 0.8cm 0.6cm, clip=true]{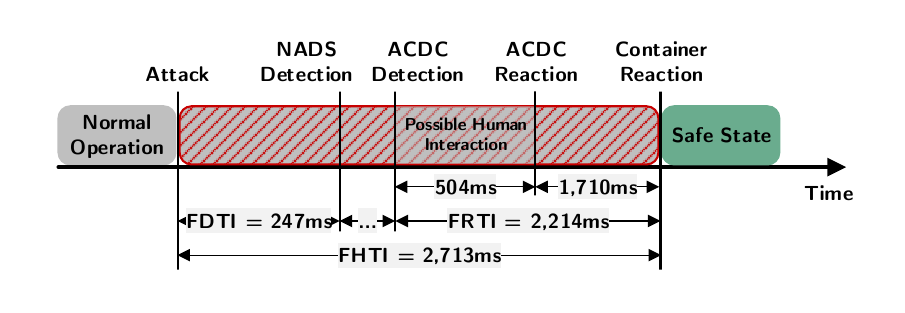}
    \caption{Maximum of cloud-based incident response times}
    \label{fig:ftti_cloud}
\end{figure}

The \ac{FHTI} for cloud-based incidence response is shown in Figure \ref{fig:ftti_cloud}. For the \ac{FDTI} we refer to the same \ac{NADS} detection time as in local incidence response. The \ac{FRTI} is composed of two sub-intervals. First, a round trip time from in-vehicle edge runtime to the \ac{ACDC} backend for an autonomous reaction. Second, the time the K3S framework takes to reallocate one application on a different \ac{HPC}. \ac{ACDC} reaction time is measured as the time between a security incident report reaching the security sensor manager and the security actuator manager receiving a response from the \ac{ACDC}. Figure \ref{fig:ftti_cloud} shows the maximum \ac{FHTI} of \SI{2713}{\milli\second} without human interaction. The measured minimum is \SI{2168}{\milli\second} and the average is \SI{2403}{\milli\second}. Compared with in-vehicle response, container reallocation results in a significantly higher \ac{FRTI} but also enables reestablishment of communication flows while the \ac{SDN} itself can only block malicious flows.

For autonomous analysis and reaction within the \ac{ACDC} Figure \ref{fig:ftti_cloud} shows the maximum sub-\ac{FRTI} of \SI{504}{\milli\second}. The measured minimum is \SI{237}{\milli\second} and the average is \SI{379}{\milli\second}. The \ac{FRTI} strongly depends on the analysis time. Manual incident and log analysis can cause higher durations. Figure \ref{fig:ftti_cloud} shows the maximum sub-\ac{FRTI} container orchestration reaction time of \SI{1710}{\milli\second}. The composition of container orchestration reaction time are shown in Table \ref{tab:co_crt}.
\begin{table}[ht]
    \centering
    \caption{Container orchestration reaction time}
    \label{tab:co_crt}
    \begin{tabularx}{\columnwidth}{l c c c c c}
        \toprule
        & \textbf{Scheduler}    & \textbf{Management}    & \textbf{Create} & \textbf{App} & \textbf{Total}\\
        \midrule
        Min & \SI{108}{\milli\second} & \SI{540}{\milli\second} & \SI{637}{\milli\second} & \SI{6}{\milli\second} & \SI{1332}{\milli\second} \\
        Avg & \SI{124}{\milli\second} & \SI{590}{\milli\second} & \SI{703}{\milli\second} & \SI{8}{\milli\second} & \SI{1426}{\milli\second} \\
        Max & \SI{157}{\milli\second} & \SI{637}{\milli\second} & \SI{975}{\milli\second} & \SI{9}{\milli\second} & \SI{1710}{\milli\second} \\
        \bottomrule
    \end{tabularx}
\end{table}

The custom scheduler module in K3S receives the desired system state, obtains the new allocation of applications and finally calls the K3S api server to schedule them accordingly. The management time describes the time interval from the K3S api server acknowledgement and the actual start of container creation. In this interval the K3S api server must distribute desired scheduling information to worker nodes and sets up networking for the POD. The create time refers to the time the container runtime takes to create the POD. Within the \ac{SOME/IP} application we measured the time it takes to register at the communication daemon.

These results are in line with the upper limits for common safety targets, which range from a few milliseconds to a few seconds.
In-vehicle countermeasures have a maximum \ac{FHTI} of \SI{328}{\milli\second} and autonomous cloud-based incident response has a maximum \ac{FHTI} of \SI{2713}{\milli\second}.
Nevertheless, our implementations leave much room for improvement: Shorter assessment intervals can improve the \ac{FDTI}, more efficient reporting mechanisms can improve the latency between detection and reaction, and container orchestration can focus on faster application start up times.

%!TEX root = ../main.tex

\section{Conclusion and Outlook}%
\label{sec:conclusion_and_outlook}

In this paper, we presented first findings from a multi-layered security infrastructure deployed in a realistic prototype vehicle and our lessons learned from the field.
The \ac{SDN} flow control reduces the \ac{IVN} attack surface by limiting the number of flows allowed in the network.
It enhances the adaptability of the network and enables incidence response strategies.
Our prototype \ac{NADS} operates with minimal false positives for control and video traffic and a feasible detection rate in our simple attack example. 
When direct countermeasures are applied to anomaly reports, however, even a very low number of false positives is too high and could lead to service disruptions in the vehicle.
Future in-car \acp{NADS} have to focus on minimizing false positives over detection rates.
Container orchestration of in-vehicle microservices enables a new incidence response strategy by moving applications between \acp{ECU}.
Nevertheless, the orchestration framework itself exhausts the resources of realistic in-vehicle hardware in our prototype implementation.
Our \ac{ACDC} implementation provides an infrastructure that is capable of monitoring and managing vehicle fleets. However, to develop anomaly detection algorithms on fleet level realistic data bases are not available in necessary quality and quantity.

In addition, we measured the time intervals for local and remote incidence response. 
With a maximum of \SI{328}{\milli\second} for in-vehicle countermeasures and \SI{2713}{\milli\second} for autonomous cloud-based incident response, our results meet the upper bounds for common safety goals, ranging from a few milliseconds to a few seconds. 
They serve as a benchmark for future incident response mechanisms in cars.

Future work shall grant further experience with our security mechanisms in practice.
Protective mechanisms for the \ac{SDN} controller and possible incidence response tailored to \acp{IVN} should be analysed for our \ac{SDN} backbone.
Different strategies, metrics and algorithms for \ac{NADS} should be compared and the filtering and fusion of reports to reduce false positives should be investigated.
A  container orchestration framework specific to the automotive domain should be developed with a reduced feature set that makes it suitable for embedded automotive systems.
Anomaly detection within an \ac{ACDC} is essential to ensure secure operation of vehicle fleets. Therefore, realistic data must be collected and anomaly detection mechanisms on fleet level must be developed.

%%%% 	BibTeX		%%%%
\bibliographystyle{IEEEtran}
\bibliography{bib/bibliography}

\end{document}